\documentclass[12pt]{JHEP3}
\pdfoutput=1
\usepackage{graphicx}
\usepackage{epsfig}
\usepackage{amssymb}
\usepackage{bm}
\usepackage{amsmath}
\makeatletter
\@addtoreset{equation}{section}
\makeatother

\title{
A Comment on  
Holographic Luttinger Theorem}
\author{
Koji Hashimoto$^{1}$ and  Norihiro Iizuka$^{2}$\\

$^1$
{\it Mathematical Physics Lab., RIKEN Nishina Center,
Saitama 351-0198, Japan}\\
E-mail: \email{koji(at)riken.jp}\\ 

$^2$
{\it Theory Division, CERN, CH-1211 Geneva 23, Switzerland}\\
E-mail: \email{norihiro.iizuka(at)cern.ch}\\
}

\abstract{
Robustness of the Luttinger theorem for fermionic liquids is examined in holography. 
The statement of the Luttinger theorem, the equality between the fermion charge density and
the volume enclosed by the Fermi surface, can be mapped to a Gauss's law in the gravity dual,
{\it a la} Sachdev. We show that various deformations in the gravity dual, such as 
inclusion of magnetic fields, a parity-violating 
$\theta$-term, dilatonic deformations, and higher-derivative corrections, do not violate the
holographic derivation of the Luttinger theorem, as long as the theory is in a confining phase.
Therefore a robustness of the theorem is found for 
strongly correlated fermions  
coupled with strongly coupled sectors which admit gravity duals.
On the other hand, in the deconfined phase, 
we also show that the deficit appearing in the Luttinger theorem
is again universal.  It measures a total deficit
which measures the charge of the  deconfined (``fractionalized'') fermions, independent of the deformation parameters.
}

\preprint{
{\normalsize CERN-PH-TH-2012-062} \\
{\normalsize RIKEN-MP-41}
}

\begin{document}
\setcounter{page}{1}

%%%%%%%%%%%%%%%%%%%%%%%%%%%%%%%%%%%%%%%%%%%%%%%%%%%%%%%%%%%%%
%%%%%%%%%%%%%%%%%%%%%%%%%%%%%%%%%%%%%%%%%%%%%%%%%%%%%%%%%%%%%

\noindent
\section{Luttinger theorem and holography}
\label{sec1}

The Luttinger theorem \cite{LM, Luttinger}
is one of the key fundamental relations in condensed matter physics and  
it states that the volume enclosed by the Fermi-surface is equal to the charge density. 
This theorem, which is originally derived by Luttinger and Ward for Landau's Fermi liquids, is  
non-trivial in the sense that this theorem is about the volume enclosed 
by the Fermi-surface. Remember that in the Landau's Fermi-liquid picture, we have a
quasi-particle description for the spectrum near the Fermi-surface, but generically the
quasi-particle description is not valid for the spectrum far away from the Fermi-surface, therefore
the spectrum deep inside the Fermi-surface does not always allow the quasi-particle description generically.  
The non-trivial point of the Luttinger theorem is that it relates the spectrum not only near the Fermi-surface 
(where quasi-particle picture holds) but also deep inside the Fermi-surface (where quasi-particle 
picture does {\it not} hold), to the charge density. 

It is widely known that the theorem holds for Fermi liquids having a Fermi surface, and 
there is a general non-perturbative proof of the Luttinger theorem for Fermi liquids \cite{Oshikawa} 
(the original proof by Luttinger and Ward was with perturbation of Fermi liquids). 
The proof of \cite{Oshikawa} is based on a $U(1)$ gauge symmetry, 
Fermi-liquids description near the Fermi-surface and a mild assumption for dynamical 
degrees of freedom, namely, all the momentum and charge carrying degrees of freedom are 
quasi-particles near the Fermi-surfaces. See also \cite{further1,further2} 
for further developments concerning 
the proof of the Luttinger theorem.

On the other hand, in nature there are quite interestingly materials, such as high $T_c$ 
superconductors or heavy fermions, 
where its normal phase shows non-Fermi liquid behavior, and in addition, the standard quasi-particle description breaks down.   
In such situations whether 
the theorem holds or not is to be better understood. 

Recent progress in applications of string theory, the holographic principle \cite{Maldacena:1997re,Witten:1998qj,Gubser:1998bc}, to condensed matter 
systems brought an insight about the Luttinger theorem in strongly correlated fermion systems.
In \cite{Hartnoll:2011dm,Hartnoll:2010xj,Iqbal:2011in}, it was pointed out that a holographic system with a charged bulk fermion,  
%in the confining geometry\footnote{This is called ``electron star" in the literature.} 
which is called ``electron star" in the literature, 
exhibits 
the Luttinger theorem of the boundary fermion theory. 
This is based on the observation that 
the bulk fermions obey a bulk Luttinger theorem at each radius for the electron 
star. 
Then, Hartnoll pointed out \cite{Hartnoll:2011fn} 
that the flux emanating from the black hole horizon will equal the 
deviation from the 
%¡Æmissing¡Ç contribution to 
%the gauge-invariant 
Luttinger relation. 
Furthermore, Sachdev  
clarified \cite{Sachdev:2011ze} 
that in a simple holographic set-up for fermions
with fermion-number chemical potential, the Luttinger relation follows simply from the Gauss's law in the bulk and 
that it holds in confined phase (thermal gas phase) but breaks down in deconfined phase (black hole phase). 
However it is also true that their argument uses a specific holographic setup like neglecting 
higher derivative corrections.    
Therefore it is natural to ask how universal the non-perturbative Luttinger theorem is 
for fermions. 
 
The holographic principle 
has been widely applied to various gravity setups, and robust correspondence has been thoroughly
studied. Among many variations of the holographic models, some of the most popular and meaningful ones
are: (i) higher-derivative corrections in the bulk gravity + Maxwell theory, 
(ii) inclusion of $\theta$ term and magnetic field,  
(iii) inclusion of a dilaton to have dilatonic gravity models.\footnote{For the inclusion 
of the dilaton, see also \cite{Huijse:2011ef,Iqbal:2011bf,Hartnoll:2011pp}.}
Each corresponds, in terms of condensed matter theory language, to: 
(i) Sub-leading terms concerning the strong coupling expansion, (ii) 
Parity-violating terms inducing quantum Hall effects under magnetic fields,
and (iii) Drastically different infra-red behavior, for example having a Lifshitz-like scaling near quantum
critical points, and more realistic systems with vanishing entropy at zero temperature.

We would like to study whether the holographic derivation of the Luttinger theorem
{\it a la } Sachdev can survive against the deformations, to find a universality of the
holographic Luttinger theorem. In this paper, we examine these popular deformations and 
show the Luttinger theorem to hold for all of these deformations, in the case of confining phases.\footnote{
The 
Luttinger theorem for various string-motivated field theories was studied in \cite{Huijse:2011hp}.}

%%%%%%%%%%%%%%%%%%%

\section{Holographic derivation of the Luttinger theorem}

We follow the beautiful argument of Sachdev's holographic derivation of the Luttinger relation \cite{Sachdev:2011ze}, 
to show the robustness of the holographic Luttinger theorem.
We generalize Sachidev's derivation \cite{Sachdev:2011ze}, in particular concerning the 
following points:
(i) In the gravity side we allow for a generic action for the bulk gauge fields including higher-derivative corrections.
(ii) We allow an axion coupling (responsible for a $\theta$-term) in the action and we include a magnetic field in the background.
(iii) The gravity action is coupled to a dilaton with a generic form of its couplings to the gravity and the
bulk gauge fields, resulting in generic background geometry which is different from the cut-off AdS$_4$ 
used in \cite{Sachdev:2011ze}. 
Specifically, we consider the following action
\begin{eqnarray}
S =&& \displaystyle\int \! d^4x
\sqrt{g}\left[R - 2 (\nabla \phi)^2-g(\phi) V_0 -(\nabla a)^2 \right.
- f(\phi) {\cal L}[F_{MN}] 
\nonumber\\
&&
\qquad\qquad
+\; i \bar{\psi} (\Gamma^M D_M + m)\psi 
\left.
- \; h(\phi)\; a \; \epsilon^{KLMN}  F_{KL}F_{MN}
\right].
\label{action}
\end{eqnarray}
Here, the spacetime dimension is 4 which is dual to a condensed matter system in 2+1 dimensions.
The $2+1$ dimensions are spanned by $x,y$ and $t$, while the $z$ direction is the emergent 
space coming out of the holographic principle. The sub-spacetime $z=0$ is the boundary of the bulk geometry.
The geometry typically is an asymptotic $AdS_4$ geometry, but in this paper we do not rely on any specific 
metric.\footnote{
The Luttinger theorem is in principle a low energy phenomena, so the UV behavior is expected not to
be relevant for the discussion.}
$D$ is the Dirac operator in which the $U(1)$ charge of the fermion is included as $q$ in the covariant derivative. 
The Maxwell field in the bulk can have a generic nonlinear electrodynamics Lagrangian 
${\cal L}[F]$. For example, the standard Maxwell Lagrangian is ${\cal L}[F]=(1/4)F_{MN}F^{MN}$, 
while the famous Dirac-Born-Infeld action which is natural in string theory
is
\begin{eqnarray}
{\cal L}(F) = \sqrt{\det \left(g_{MN}+ \frac{1}{\lambda} F_{MN}\right)},
\end{eqnarray}
and includes higher derivative terms as a form of $F^4$ and higher multiples.
In holography, higher-derivative corrections may be related directly to a physical consequence; 
for example, the famous calculation on the shear viscosity of the quark-gluon plasma 
\cite{Policastro:2001yc,Kovtun:2004de}
can be corrected by higher-derivative terms to have lower values, but a physical constraint on
the form of the higher-derivative terms coming from a bulk causality 
may forbid the value of the viscosity (divided by the entropy density) 
to go lower \cite{Brigante:2008gz,Brigante:2007nu}. 
In generic holographic setting, even if we take large $N$ limit, these higher derivative corrections 
are non-negligible. 

The system couples to the bulk scalar field $\phi$ which is a dilaton in string theory.
Explicit solutions (such as the ones with horizons studied in, for examples,  
\cite{Gubser:2009qt} - \cite{Gouteraux:2011qh}
) are not necessary in the following.
In the second line of the action \eqref{action}, we have the parity-violating
$\theta$-term (the axion coupling), which is relevant for a quantum Hall effect under the magnetic field,  
see for examples, \cite{KeskiVakkuri:2008eb,Fujita:2009kw,Bak:2009kz,Goldstein:2010aw,Bayntun:2010nx}.

The essence of the Sachdev's derivation
is to consider a bulk fermion one-loop path-integration in the total free energy
as a semi-classical approximation. 
This affects the scalar potential of the bulk gauge field, to minimize the free energy (which is nothing but the on-shell 
effective action in the gravity side). 
A radial integral of the bulk Gauss's law turns out to be nothing
but the Luttinger relation. 
Once one puts $\phi=a=0$ and ${\cal L} = (1/4) F^2$, and takes a cut-off AdS space in the derivation below,
everything reduces to the Sachdev's original derivation. 

The bulk free energy per a unit volume in our case is
\begin{eqnarray}
&&{\cal F} = \int \! dz \; \sqrt{g} 
\left(-f(\phi) {\cal L}[F] - h(\phi) a F \tilde{F}\right)
- \frac{T}{V} {\rm Tr} \;{\rm Log}\; \left[D\cdot \Gamma + m\right].
\label{freeene}
\end{eqnarray}
We have integrated out the bulk fermion $\psi$, to have the last term.\footnote{If we do not include this
fermion path integral to account for the back-reaction to the gauge potential, we would not obtain
the Luttinger relation (see for example a discussion in \cite{Larsen:2010jt}). 
The effects of the bulk fermions should be communicated
with the gauge potential, as the Luttinger theorem is a relation between the Fermi surface and the charge density.}
We ignore the back reaction of the fermions to the bulk geometry and the dilaton and the axion; we only consider a back reaction
to the gauge field (a possible justification is presented in the next section).

As for the gauge field configuration, we assume the homogeneity and the isotropy in the $(x,y)$ directions. Then  
non-vanishing components of the static gauge field strengths are only $F_{tz}$ and $F_{xy}$. Using the Jacobi identity
in the $(x,y,z)$ space, we obtain $\partial_z F_{xy}=0$ which means that $B\equiv F_{xy}$ is constant. So, in the
$A_z=0$ gauge, we are left with the gauge configuration
\begin{eqnarray}
A_0=A_0 (z), \quad A_x=-\frac{B}{2}y,\quad  A_y=\frac{B}{2}x, \quad A_z=0.
\label{gaugef}
\end{eqnarray}

The charge density of the boundary theory is given by 
\begin{eqnarray}
\langle {\cal Q} \rangle
\equiv -\frac{\partial {\cal F}}{\partial \mu}
=-D_z(z=0)
%\end{eqnarray}
\quad
\mbox{where}
\quad 
%\begin{eqnarray}
D_z \equiv 
f(\phi) \frac{\partial {\cal L}[F]}{\partial F_{0z}} + h(\phi) a F_{xy}.
\label{defch}
\end{eqnarray}
Here we have defined the chemical potential as $\mu = \Phi(z=0)$ where we denote the temporal component 
of the gauge field as $A_0\equiv i \Phi$. 
Note that the $\Phi$ dependence in the fermion loop in \eqref{freeene} does not contribute
to the definition of the charge \eqref{defch}, 
because the bulk fermion wave functions vanish at the boundary $z=0$ where the chemical potential
is defined, due to the normalizability.

To explicitly perform the fermion one-loop integral to evaluate \eqref{freeene}, 
one just needs a formal expression characterizing the
discrete energy eigen modes of the bulk fermions, which looks
\begin{eqnarray}
{\cal D}_n \; \chi_{l,n}(z) = E_{l,n}\;\chi_{l,n}(z).
\label{eigen}
\end{eqnarray}
The operator ${\cal D}_n$ is nothing but a spatial part of the covariant Dirac operator in the curved geometry.
$\chi_{l,n}$ and $E_{l,n}$ are the eigenfunctions and eigenvalues of the operator ${\cal D}_n$, respectively.\footnote{
For deconfined geometry $l$ can be a continuous parameter, but here we formally write the generic 
label as $l$. In addition to that, in the deconfined geometry, 
the energy eigenvalues are generically complex,
thus the amplitude of the wave function damps in time exponentially, as in the case of quasi-normal modes. Here formally we regard our calculation performed in a time scale shorter than the decay
time scale.}
Here the integer $l$ labels the Kaluza-Klein modes of the bulk fermion in the curved space, and $n$ labels
the Landau levels of the fermion wave function in the $x$-$y$ space as the magnetic field is present.
Below, we shall use only the two facts: First, the operator ${\cal D}_n$ includes a trivial term $q  \, \Phi$ coming from the
minimal coupling in the Dirac operator, and second, the normalization is given as
\begin{eqnarray}
\int \! dz \sqrt{\frac{g_{zz}}{-g_{tt}}} \; \chi_{l,n}^\dagger (z) \chi_{l,n}(z) = 1 \,,
\label{norm}
\end{eqnarray}
where no summation for $l$ and $n$ is imposed.\footnote{We consider the cases where bulk metric takes the form as $ds^2 =g_{tt}(z) dt^2 + \sum_{i=1}^2 g_{ii}(z) dx_i^2 + g_{zz}(z) dz^2$.} See Appendix A for the explicit evaluation of the states with the operator ${\cal D}$.

Using this energy eigenvalue, generically the trace log term in the free energy can be evaluated as 
\begin{eqnarray}
\frac{T}{V}\;{\rm Tr} \;{\rm Log}\; \left[D\cdot \Gamma + m\right]
= \frac{qB}{2\pi} \sum_{l}\sum_n E_{l,n} \theta(-E_{l,n}).
\label{fermione}
\end{eqnarray}
The step function is necessary to count only the residues appearing in the shift of
the poles in 
the path integral in the off-shell $k^0$ space. See Appendix B for the detailed calculations.
The factor $qB/2\pi$ is the unit volume of the discretized momentum space $(k_x,k_y)$ due to the magnetic field.

Now, to find a saddle point of the free energy with respect to the bulk field $\Phi$, we consider the bulk on-shell equation 
\begin{eqnarray}
0 = \frac{\delta {\cal F}}{\delta \Phi}.
\label{df}
\end{eqnarray}
To calculate this with the fermion loop term, we use \eqref{norm} and \eqref{eigen} to rewrite the fermion 
free energy \eqref{fermione} as
\begin{eqnarray}
E_{l,n} \theta(-E_{l,n} ) 
&=& \int \! dz \sqrt{\frac{g_{zz}}{-g_{tt}}} \; \chi^\dagger_{l,n} (z) \chi_{l,n}(z) E_{l,n}  \theta(-E_{l,n}) \nonumber \\
&=&  \int \! dz \sqrt{\frac{g_{zz}}{-g_{tt}}} \; \chi^\dagger_{l,n} (z) {\cal D} \chi_{l,n}(z) \theta(-E_{l,n}) \,.
\end{eqnarray}
With the fact that $\Phi$ dependence of the operator 
${\cal D}$ is just linear in $q \Phi$ as it is a Dirac operator (see Appendix A for the details), 
the minimization of the free energy \eqref{df} is\footnote{When taking a variation of $E \, \theta(-E)$ 
with respect to $\Phi$, 
one may be worried about the $\Phi$-dependence in $E$ inside the step function. However, as the variation of $\theta(E)$
is a delta function while there is an overall $E$ in front of it, the worring contribution disappears.}
\begin{eqnarray}
-
\partial_z
\left(
f(\phi) \frac{\partial {\cal L}[F]}{\partial F_{0z}} + h(\phi) a F_{xy}
\right)
 -q \frac{qB}{2\pi} \sqrt{\frac{g_{zz}}{-g_{tt}}} \sum_{l,n} \theta(-E_{l,n}) \chi^\dagger_{l,n}(z) \chi_{l,n}(z) = 0 \,. \quad \quad
 \label{l2}
\end{eqnarray}
Then we make an integration over the $z$ space, which leads us to the Luttinger relation, as follows.
For that, we need one more information for the geometry at the IR.
Let us first consider a generic confining geometry, for which the geometry consistently ends at $z=z_{\rm IR}$.
The Gauss's law at the IR end $z=z_{\rm IR}$ shows that the electric flux $D_z$ vanishes
there. Then, from the $z$ integration of \eqref{l2}, we obtain
\begin{eqnarray}
\langle {\cal Q}  \rangle /q =  \frac{qB}{2\pi} \sum_{l,n} \theta(-E_{l,n}) \,.
\label{l}
\end{eqnarray}
This is the Luttinger relation, since the right hand side is the volume enclosed by the 
Fermi surface. Note that since we turn on the magnetic field, the Landau levels appear and the $x$-$y$ momentum
space is discretized, and resultantly the unit volume of the $x$-$y$ momentum space $qB/2\pi$ appears.

For deconfined geometries, the IR boundary condition differs, and in particular the electric flux
does not vanish there.
There appears a deficit in the Luttinger relation. We discuss
the situation in the next section.

One should have noticed that the derivation here is almost identical with what Sachdev gave in
\cite{Sachdev:2011ze}. However, we find it intriguing that the derivation by Sachdev is so robust
that the theorem is valid against various deformations of the system. In particular, the inclusion of 
the  higher derivative corrections corresponds  
to the direction toward a weak coupling where
it is plausible that the Luttinger theorem is valid. Furthermore, the inclusion of the background
magnetic field is interesting, as it not only introduces a nice regularization of the momentum space
but also is involved with quantum Hall effects. The dilatonic corrections are related with existence 
of different scaling at the IR, and even in those systems our generalized derivation shows that the
Luttinger theorem holds.

Although in this paper we worked in four spacetime dimensions in the gravity theory (which corresponds to
three spacetime dimensions for the liquid system), we can generalize the derivation to higher dimensions, in a 
straightforward manner. A possible obstacle would be the $F\wedge F$ term in the bulk, which should be generalized to 
a Chern-Simons term in higher dimensions, then one cannot impose the isotropy which we have employed in 
the derivation above. Another concern may be on the fermion integral, since generically in higher dimensions
the theory becomes non-renormalizable. However, the leading quantum loop which we considered in this 
paper is just an effect of the chemical potential of the one-loop diagram of fermions, 
which can be defined without any problem. Therefore we 
claim that higher-dimensional generalization of the derivation can be done accordingly.

%%%%%%%%%%%%%%%%%%%%%%%%%%%%%%

\section{Discussions}

The essential statement which we would like to make in this paper is just the robustness 
of the holographic Luttinger theorem,
which was already shown in the previous section. We end this short paper with 
two important observations: First, the validity of our calculation in view of quantum
corrections in string theory and AdS/CFT, and second, the emergent dependence on
deformation parameters only in the deconfinement phase.

\subsection{Quantum corrections in string theory}
In the derivation of the holographic Luttinger theorem, Sachdev and we computed 
the fermion one-loop diagram and considered its back-reaction to the gauge potential.
However, in general in string theory, other loops involving gravity and other fields
may contribute, so the effect on the gauge potential may not be only from
the fermion one-loop. Here we shall point out that a proper large $N$ scaling 
in AdS/CFT can avoid this mixing problem\footnote{See also \cite{Sachdev:2011ze} for the argument 
to include quantum fluctuations beyond the fermion one-loop.}.

We assume that the bulk fermion is from a space-filling D-brane. This fermion is
often called ``mesino" since it may be a fermion counterpart of mesons,
when the D-brane is identified with the flavor D-branes in the AdS/CFT correspondence 
\cite{Karch:2002sh}. 
The gauge potential $A_\mu$ also comes from a
space-filling D-brane. This means that the effective gauge 
coupling $q$ is of order 
${\cal O} (1/\sqrt{N})$, because in the AdS/CFT correspondence, the string
coupling constant $g_s$ scales as $\sim 1/N$, and the coupling on the 
D-brane is the open string coupling $\sqrt{g_s}$ while the coupling in the bulk 
geometry (gravity and the dilaton $\phi$ and the axion $a$) is $g_s$.
Denoting the graviton/dilaton/axion fluctuation as $\delta g$, and
the gauge fluctuation as $\delta A_\mu$, and the fermion fluctuation 
as $\psi$, then the generic dependence in $N$ in general AdS/CFT is written as
\begin{eqnarray}
S &=& (\partial \delta g)^2 + \frac{1}{N} (\delta g)^3 + \frac{1}{N^2} (\delta g)^4
\nonumber \\
&+& (\partial \delta A_M)^2 + 
\bar{\psi}\Gamma^M D_M \psi +
 \frac{1}{\sqrt{N}} (\delta A_M) \bar{\psi}\psi
 + \frac{1}{N} \delta g (\delta A_M)^2
 + \frac{1}{N} \delta g \bar{\psi}\psi + \cdots.
\end{eqnarray}
Looking back our derivation of the holographic Luttinger theorem,
we have performed the $\psi$ integral in this action. From the generic
action written above, we observe that this one-loop integral involves the term
$\frac{1}{\sqrt{N}}(\delta A_\mu)\bar{\psi}\psi$, so it shifts the action by a term
of order 
${\cal O}(1/\sqrt{N})$. This is a leading order effect, compared to the
quantum corrections involving the graviton, the dilaton and the axion,\footnote{Here, as a classical background geometry,
we have assumed that
the back-reaction of the flavor brane itself (with $\psi=0$) is already included in the geometry. 
In this paper we need not to
specify the geometry for the derivation of the Luttinger theorem, so whatever the back-reacted classical geometry is,
there is no problem in the derivation.} since
in the action above those corrections come with a coupling of ${\cal O}(1/N)$.

Therefore, we conclude that, assuming that 
the origin of the fermions and the gauge fields is D-branes in the bulk,
the fermion one-loop integral is the leading order in $1/N$ expansion in
AdS/CFT, so we can consistently ignore the other quantum corrections.

\subsection{Deconfined phase}
As first pointed out by Hartnoll \cite{Hartnoll:2011fn} and also 
studied in detail by Iqbal and Liu recently \cite{Iqbal:2011bf}, the presence of the
black hole horizon significantly alters the result; in the final line of the derivation of the
Luttinger theorem \eqref{l}, we have used the fact that the electric displacement
$D_z$ at the IR endpoint vanishes due to the confining geometry.
However in the presence of the black hole the IR boundary condition is different
and there exists in general an electric flux emanating from the black hole
horizon, for charged black holes.
\begin{eqnarray}
\langle {\cal Q}\rangle + D_{z}\!\bigm|_{z=z_{\rm IR}}
= q \sum_{l,n}\theta (-E_{l,n}) \,,
\label{ludef}
\end{eqnarray}
with 
\begin{eqnarray}
\label{electricdisplacement}
D_z \!\bigm|_{z=z_{\rm IR}} \equiv  \left(f(\phi) \frac{{\cal{L}}[F]}{\partial F_{0z}} + h(\phi) a F_{xy} 
\right) \!\bigm|_{z=z_{\rm IR}} \,.
\end{eqnarray}
So, there exists a Luttinger deficit for the deconfined phase \cite{Hartnoll:2011fn, Iqbal:2011bf}.
Here $D_z$ is the electric displacement in the bulk at the IR of the geometry,
which is the
electric flux penetrating the horizon of the black hole. This 
comes from the first term of \eqref{l2}.

In the presence of the deformations which we consider, for the deconfined phase,
there appears a dependence on the magnetic field and the axion field
at the horizon. The axion field corresponds to the parity-violating 
$\theta$ term, so in general, the Luttinger theorem in the deconfinement phase
is violated with a deficit dependent on the $\theta$ term and the magnetic field
in addition to the deficit \cite{Iqbal:2011bf} 
of the electric charge carried by the ``fractionalized" fermions \cite{Senthil:2003,Sachdev:2010um,Huijse:2011hp} 
(which are deconfined quarks in the standard holographic QCD terminology).

The electric displacement includes a contribution from the 
parity-violating term,
the second term of the (\ref{electricdisplacement}),  
\begin{eqnarray}
\left[D_z\!\bigm|_{z=z_{\rm IR}}\right]_{\rm parity-odd}
= \theta_{\rm eff} B.
\end{eqnarray}
This term directly responds to the magnetic field present in the system, and the effective 
value of the coefficient, $\theta_{\rm eff}$, in the Luttinger theorem is given by
\begin{eqnarray}
\theta_{\rm eff} \equiv [h(\phi)a]_{z=z_{\rm IR}}.
\end{eqnarray}
It is intriguing that the parameter $\theta_{\rm eff}$ is not given by the UV geometry but the
IR geometry. The IR geometry is not directly related to the parameters of the fermion liquid system 
defined at UV, and it is rather determined by the gravitational dynamics corresponding to strongly 
coupled sectors, which these fractionalized fermions couple.  
This dependence on IR geometry reflects the fact that the 
Luttinger relation is a phenomena at low energy.
Note that even with no electric field $F_{0z}$, once the parity-violating term and the magnetic field 
$F_{xy}$
is turned on, the Luttinger deficit appears.

The deficit appearing in the Luttinger relation \eqref{ludef} appears to depend explicitly on
the deformation parameters we introduced in the gravity dual. The parameters
are in the nonlinear electrodynamics ${\cal L}[F]$ and the dilaton-axion couplings
to the gauge fields in the gravity side. However, in \eqref{ludef}, the deficit depend
only on the electric displacement $D_z$ at the black hole horizon. The effect
of the parity-odd term ($\theta_{\rm eff} B$) is also included in the
electric displacement. So, we conclude that the effect of the
deconfinement phase to the deviation from Luttinger theorem 
can be summarized into electric displacement $D_z$, which is determined by 
the total charge of the ``fractionalized" fermions, even in the presence of the deformations considered in this paper. Note that we have not assumed the existence of the quasi-particle picture in our derivation.

%%%%%%%%%%%%%%%%%%%%%%%%%%%%%%%%%%%%%%%%%%%%%%%

\acknowledgments
We would like to especially thank Masaki Oshikawa for his very nice lecture at RIKEN. 
K.H.~is partly supported by
the Japan Ministry of Education, Culture, Sports, Science and
Technology. N.I.~would like to thank Mathematical physics laboratory in RIKEN for 
very kind hospitality.

%%%%%%%%%%%%%%%%%%%%%%%%%%%%%%

%%%%%%%%%%%%%%%%%%%%%%%%%%%%%%%%%%%%%%%%%%%

\appendix

\section{Bulk fermion eigen-states}

In this appendix we shall explicitly calculate \eqref{eigen} with the background gauge field.
We work in the Lorentzian  signature, while the Euclidean signature (which we employed in
the derivation) can be easily obtained by an analytic continuation.
First, we derive the Dirac operator. The fermion action in the bulk is
\begin{eqnarray}
S_{\rm fermion} = \int \! d^{3+1}x \; \sqrt{-g} \, i \,
\left[\, \bar{\psi} \Gamma^M D_M \psi - m\bar{\psi}\psi \, \right].
\end{eqnarray}
Here the Dirac operator is $D_M = \partial_M + \frac{1}{4} w_{abM}\Gamma^{ab} - i q A_M$. 
The definition of the
Gamma matrices in the local Lorentz frame are
\begin{eqnarray}
\Gamma^{\underline{z}} \equiv \left(
\begin{array}{cc}
{\bf 1}_2 & {\bf 0}_2 \\
{\bf 0}_2 & {\bf -1}_2
\end{array}
\right),
\quad
\Gamma^{\underline{\mu}} \equiv \left(
\begin{array}{cc}
{\bf 0}_2 & \gamma^\mu \\
\gamma^\mu & {\bf 0}_2
\end{array}
\right)
\end{eqnarray}
with $\gamma^0 \equiv i\sigma_3$, $\gamma^1 \equiv \sigma_1$, and $\gamma^2\equiv -\sigma_2$, 
where $\sigma_1, \sigma_2, \sigma_3$
are the Pauli matrices. We follow the notation of 
\cite{Liu:2009dm} and \cite{Iizuka:2011hg} except for the assignment of $\gamma^\mu$
(this difference is necessary to see the diagonalization as for fermion components, see below).
The notation for the indices are: $M=0,1,2,z$, and $\mu=0,1,2$.

Writing the 4-component fermion as 
\begin{eqnarray}
\psi \equiv \left(
\begin{array}{c}
\psi_+
\\
\psi_-
\end{array}
\right), \quad 
\psi_\pm \equiv (-g g^{zz})^{-1/4} \phi_{\pm},
\end{eqnarray}
where $\phi_\pm$ is a two-spinor,
the Dirac equation is 
\begin{eqnarray}
\sqrt{\frac{g_{ii}}{g_{zz}}} (\partial_z \mp m \sqrt{g_{zz}})\phi_\pm
= \mp i K_\mu \gamma^\mu \phi_{\mp},
\label{dirack}
\end{eqnarray}
with $K_0 \equiv -i \sqrt{\frac{g_{ii}}{-g_{tt}}} (\partial_0 - i q A_0)$ and $K_i \equiv -i (\partial_i - i q A_i)$ 
with $i=1,2$.

In the Dirac equation, the four spinor components are coupled, while we would like to group them into
2-spinors to derive \eqref{eigen}. In our convention, among the gamma matrices, $\gamma^0$ is
a diagonal matrix while $\gamma^i$ is not, so if we can bring the $K_i \gamma^i$ to a diagonal form, 
the decomposition to the 2-spinors is complete. This is nothing but solving the following 
eigen equation in the $x$-$y$ space,
\begin{eqnarray}
i(K_1 \gamma^1 + K_2 \gamma^2) 
\tilde \phi_{\pm}
= 
\left(
\begin{array}{cc}
\alpha & 0 \\ 0 & \beta
\end{array}
\right)
\tilde \phi_{\pm},
\label{eigeneq2}
\end{eqnarray}
where $\alpha$ and $\beta$ are complex constants, 
and $\tilde{\phi}_\pm (x,y)$ are 2-spinor wave functions representing 
Landau levels of fermions in magnetic fields. The gauge field configuration \eqref{gaugef} satisfies
$[K_1,K_2]=  i q B$, so using a creation and an annihilation operator
\begin{eqnarray}
\frac{1}{\sqrt{2qB}}(K_1+ iK_2) \equiv a, \quad
\frac{1}{\sqrt{2qB}}(K_1- iK_2) \equiv a^\dagger,
\end{eqnarray}
we have the harmonic osccillator $[a,a^\dagger]=1$. The eigen equation \eqref{eigeneq2} becomes
\begin{eqnarray}
i\sqrt{2qB}
\left(
\begin{array}{cc}
0 & a \\
a^\dagger & 0
\end{array}
\right)
\tilde \phi_{\pm}
= 
\left(
\begin{array}{cc}
\alpha & 0 \\ 0 & \beta
\end{array}
\right)
\tilde \phi_{\pm} .
\end{eqnarray}
A solution can be easily found as
\begin{eqnarray}
\tilde \phi_{\pm} 
\propto \left(
\begin{array}{c}
(i  \sqrt{2qBn}/\alpha) (a^\dagger)^{n-1}
|0\rangle
\\
(a^\dagger)^n
|0 \rangle
\end{array}
\right) \,, \quad n=0,1,2,\cdots.
\end{eqnarray}
We need $\beta= -2qBn/\alpha$ to have this solution.
With this wave function for $(x,y)$, 
we choose 
\begin{eqnarray}
\phi_{\pm,n} = f_{\pm,n} (z) e^{-iw_\pm t}\left(
\begin{array}{c}
(i  \sqrt{2qBn}/\alpha_{\pm})
 (a^\dagger)^{n-1}
|0\rangle
\\
(a^\dagger)^n
|0 \rangle
\end{array}
\right) \,,
\end{eqnarray}
then  
the Dirac equation \eqref{dirack} is written as
\begin{eqnarray}
\label{almostfinalDiraceq}
&& \sqrt{\frac{-g_{tt}}{g_{zz}}}  \left(\partial_z \mp {m}\sqrt{g_{zz}}\right) \phi_{\pm,n} \nonumber \\
&&  = \mp \left(
(i\partial_t + q A_0)\sigma_3
+
\sqrt{\frac{- g_{tt}}{g_{ii}}}
\left(
\begin{array}{cc}
\alpha_{\mp} & 0 \\ 0 & -2qBn/\alpha_{\mp}
\end{array}
\right)
\right)\phi_{\mp ,n} \,. 
\end{eqnarray}

From the phase matching in (\ref{almostfinalDiraceq}), we have $w_+=w_- \equiv E_n$. 
The upper spinor component of the equation (\ref{almostfinalDiraceq}) yields
\begin{eqnarray}
&&  
\label{almostfinaldiraceqforone}
\sqrt{\frac{-g_{tt}}{g_{zz}}}  \left(\partial_z - {m}\sqrt{g_{zz}}\right)
 \frac{f_{+,n}(z)}{\alpha_+} 
= -\left(E_n + q A_0 +\sqrt{\frac{- g_{tt}}{g_{ii}}}\alpha_-
\right)  \frac{f_{-,n}(z)}{\alpha_-} \,, \quad \quad
\\
&&
\label{almostfinaldiraceqfortwo}
\sqrt{\frac{-g_{tt}}{g_{zz}}}  \left(\partial_z + {m}\sqrt{g_{zz}}\right)
 \frac{f_{-,n}(z)}{\alpha_-} 
=  \left(E_n + q A_0 + \sqrt{\frac{- g_{tt}}{g_{ii}}}\alpha_+
\right) \frac{f_{+,n}(z)}{\alpha_+} \,.
\end{eqnarray}
The lower spinor component of the equation (\ref{almostfinalDiraceq}) yields 
\begin{eqnarray}
&&  
\label{almostfinaldiraceqforone1}
\sqrt{\frac{-g_{tt}}{g_{zz}}}  \left(\partial_z - {m}\sqrt{g_{zz}}\right)
f_{+,n}(z)
= -\left(-(E_n + q A_0) +\sqrt{\frac{- g_{tt}}{g_{ii}}}(-2qBn/\alpha_-)
\right) f_{-,n}(z) \,, 
\nonumber
\\ \\
&&
\label{almostfinaldiraceqfortwo2}
\sqrt{\frac{-g_{tt}}{g_{zz}}}  \left(\partial_z + {m}\sqrt{g_{zz}}\right)
f_{-,n}(z)
=  \left(-(E_n + q A_0) + \sqrt{\frac{- g_{tt}}{g_{ii}}}(-2qBn/\alpha_+)
\right) f_{+,n}(z) \,.
\nonumber \\
\end{eqnarray}
For (\ref{almostfinaldiraceqforone1}) and (\ref{almostfinaldiraceqfortwo2}) to be consistent with (\ref{almostfinaldiraceqforone}) and (\ref{almostfinaldiraceqfortwo}),
we need
\begin{eqnarray}
{\alpha_+}=-{\alpha_-}, \quad \alpha_-=2qBn/\alpha_-, \quad \alpha_+ = 2qBn/\alpha_+.
\end{eqnarray}
This can be solved as
\begin{eqnarray}
\label{twoindependentsolutions}
\alpha_\pm= \pm \sqrt{2qBn} \quad {\rm or}\quad
\alpha_\pm= \mp\sqrt{2qBn} \,.
\end{eqnarray}
Then (\ref{almostfinaldiraceqforone1}) and (\ref{almostfinaldiraceqfortwo2}) become the same as 
(\ref{almostfinaldiraceqforone}) and (\ref{almostfinaldiraceqfortwo}), 
and we obtain two independent solutions as (\ref{twoindependentsolutions}).

The resultant equations 
(\ref{almostfinaldiraceqforone1}) and (\ref{almostfinaldiraceqfortwo2})
can be re-written as
\begin{eqnarray}
&& \quad \quad \quad \quad \quad \quad \quad \quad \quad {\cal D}_n^{(i=1)} \; 
\chi_n(z) = E_n \; \chi_n(z)  \,, 
\\
&&{\cal D}^{(i=1)}_n \equiv \sqrt{\frac{-g_{tt}}{g_{zz}}}\left(
-i \sigma_2 \partial_z - \sigma_1 {m}{\sqrt{g_{zz}}} \right) - q A_0 -  \sigma_3 \sqrt{\frac{- g_{tt}}{g_{ii}}} 
\sqrt{2qBn} 
 \,, \quad \quad
 \label{dn1} 
\end{eqnarray}
with $\chi \equiv (f_+,f_-)^{\rm T}$. This is for the choice $\alpha_{\pm}=\pm \sqrt{2qBn}$.
For the other choice $\alpha_{\pm}=\mp\sqrt{2qBn}$,
\begin{eqnarray}
{\cal D}^{(i=2)}_n \equiv \sqrt{\frac{-g_{tt}}{g_{zz}}}\left(
-i \sigma_2 \partial_z - \sigma_1 {m}{\sqrt{g_{zz}}} \right) - q A_0 +  \sigma_3 \sqrt{\frac{- g_{tt}}{g_{ii}}} 
\sqrt{2qBn} \,. 
\label{dn2}
\end{eqnarray}

The operator ${\cal D}_n$ of \eqref{eigen}
appears here as ${\cal D}_n^{(i)}$. See also \cite{Sachdev:2011ze,Huijse:2011ef}.
Note that due to the interaction term between spins and magnetic fields, 
there are two ${\cal D}_n$'s labeled by $i$.  
Generically this
eigen-equation allows only a discrete spectrum\footnote{
For given index $n$ and $i$, the corresponding parts of Dirac equation 
be coupled 1st order differential equations 
for two spinor components, 
so it allows two independent solutions. 
Taking the appropriate ration between these two, we can eliminate non-normalizable mode in the 
bulk UV and regular mode in bulk IR in Euclidian signature. Then it allows only discrete modes 
labeled by $l$.} 
for $E_n$, so the energy $E_n$
is also labeled by $l$ (the Kaluza-Klein modes in $z$ space) in addition to $n$ (the Landau levels) 
and $i$ (spins) as 
\begin{eqnarray}
\label{onedimdiraceq}
{\cal D}^{i}_n \; 
\chi_{n,l,i}(z) = E_{n,l,i} \; \chi_{n,l,i}(z)  \,.
\end{eqnarray}
For simplicity of the notation, we will omit the index $i$ in this paper.

The normalizability condition should single out a certain linear combination of the two solutions.
The reason why we got two solutions generically 
can be easily understood by the following argument. The Dirac
equation \eqref{dirack} is a coupled equation of $\phi_+$ and $\phi_-$, but one can eliminate
one of them. Bringing \eqref{dirack} into the following form formally,
\begin{eqnarray}
( - i K_\mu \gamma^\mu)^{-1}\sqrt{\frac{g_{ii}}{g_{zz}}} (\partial_z - m \sqrt{g_{zz}})\phi_+
= \phi_-,
\label{phi-so}
\\
( i K_\mu \gamma^\mu)^{-1}
\sqrt{\frac{g_{ii}}{g_{zz}}} (\partial_z + m \sqrt{g_{zz}})\phi_-
=  \phi_+,
\end{eqnarray}
we can combine these to eliminate $\phi_-$, to have
\begin{eqnarray}
\left[
( i K_\mu \gamma^\mu)^{-1}
\sqrt{\frac{g_{ii}}{g_{zz}}} (\partial_z + m \sqrt{g_{zz}})
( - i K_\mu \gamma^\mu)^{-1}\sqrt{\frac{g_{ii}}{g_{zz}}} (\partial_z - m \sqrt{g_{zz}})
\right]\phi_+
= \phi_+.
\end{eqnarray}
This is a 2nd order differential equation for a two-spinor $\phi_-$, 
so we generically have four independent solutions. Without magnetic field, 
given $E$, four states are degenerate and they correspond to 
spin \{up and down\}, and \{normalizable and non-normalizable\} modes. 
In the presence of magnetic fields, 
this degeneracy splits up by spins, so given spin and energy eigenvalue $E$, 
there is a set of a normalizable and a non-normalizable mode.

%%%%%%%%%%%%%%%%%%%

\section{Evaluation of the fermion free energy}

In this appendix, we demonstrate the calculation of the fermion one-loop free energy
\eqref{fermione}.

\subsection{Reduction to $1+0$ dimension}

In the previous appendix, we demonstrate that given 
$l$ (the Kaluza-Klein modes in $z$ space), $n$ (the Landau levels) 
and $i$ (spins),  the Dirac equation reduces to a single
equation (\ref{onedimdiraceq}) which determines $E_{l,n}$. Therefore the action for the fermion should
read as
\begin{eqnarray}
S = \sum_{l,n} \int \! dt \; i \bar{\Psi}_{l,n}(t) \left(i\partial_t - E_{l,n}\right) \Psi_{l,n}(t).
\end{eqnarray}
Here $\Psi_{l,n}(t)$ is a single component fermion, since the dependence on the Landau level $n$ is
already included with the specific Landau level wave function in the $x$-$y$ space.
The fermion field $\Psi_{l,n}(t)$ is properly normalized to have the action above.
Note also that there are two $E_{l,n}$'s, depending on the choice of the Dirac operator ${\cal D}_n$ 
in \eqref{dn1} and \eqref{dn2} labeled by $i$. 
In the following, we omit the index $i$  for simplicity of the notation. 

Let us proceed to calculate the free energy.
We bring the Tr Log into the following expression,
\begin{eqnarray}
{\cal F}_{\rm fermion}
&\equiv& \sum_n \frac{qB}{2\pi} 
\sum_l {\rm Tr} \;{\rm Log} \left(i\partial_t - E_{l,n}\right) 
= \frac{qB}{2\pi} 
\sum_{l,n} \int^\infty_{E_{l,n}} \! ds \; {\rm Tr}\frac{1}{ \left(i\partial_t - s\right) } \nonumber
\\
&=&
\frac{qB}{2\pi}\sum_{l,n} \int^\infty_{E_{l,n}} \! \! ds
\int \!\frac{dw}{2\pi}
\frac{1}{w-s} \,.  
\end{eqnarray}
Here, 
we have used the fact that the momentum integration in the $k_x$-$k_y$ space
is now replaced by the Landau level summation, as
\begin{eqnarray}
\int \frac{dk_xdk_y}{(2\pi)^2} =\int \frac{2 \pi k dk}{(2\pi)^2} = \int \frac{\pi d(k^2)}{(2\pi)^2} =
\sum_n  \frac{2\pi qB}{(2\pi)^2} = \sum_n \frac{qB}{2\pi} \,.
\end{eqnarray} 
We have used the momentum relation $k^2= 2 q B n$ which we obtained in the last appendix.

Now we perform the off-shell $w$ integration. 
The standard path in the complex $w$-plane for the
integration of $w$ rounds the upper half plane plus the real axis.
In the $i\epsilon$ prescription,
the pole contributing in the path integral is the one on the negative real 
axis of $w$. This appears only for the negative $s$   
so the integral is non-zero only when $E_{l,n}<0$, 
and we obtain 
\begin{eqnarray}
{\cal F}_{\rm fermion} 
= \frac{qB}{2\pi} \sum_{l,n} \int_{E_{l,n}}^0 \! ds \frac{1}{2\pi} 2\pi i \theta(-E_{l,n}) 
= -i \frac{qB}{2\pi} \sum_{l,n} E_{l,n} \theta(-E_{l,n}).
\end{eqnarray}
This expression is \eqref{fermione} in the Euclidean notation.

\subsection{Another viewpoint: Reduction to $1+2$ dimension}

We shall present another viewpoint here, to evaluate the fermion free energy, which is
a natural dimensional reduction along $z$.

First, we shall decompose the bulk fermion into Kaluza-Klein modes along $z$ 
labeled by $l$. The effective action for the $l$-th mode has a kinetic operator
$D_\mu \gamma^\mu - m_l$ where $m_l$ is the mass for the $l$-th mode of the
decomposed fermion.
Then, we compute ${\rm Tr} \;{\rm Log} (D_\mu \gamma^\mu - m_l)$ for each state $l$
and make a summation over $l$. 

Let us work out the Kaluza-Klein decomposition explicitly.
The bulk fermion action is 
\begin{eqnarray}
\label{B54dimaction}
S_{\rm fermion} = \int \! d^{3+1}x \; \sqrt{-g} \, i \,
\left[\, \bar{\psi} \Gamma^M D_M \psi - m\bar{\psi}\psi \, \right].
\end{eqnarray}
Using the notation in the previous appendix, this action $S_{\rm fermion}$ 
can be explicitly written as
\begin{eqnarray}
\int \! d^{3+1}x \; \sqrt{\frac{g_{zz}}{- g_{tt}}} \, i \,
\left(\phi_+^\dagger \gamma^0, \phi_-^\dagger \gamma^0\right)
\left(
\begin{array}{cc}
i \bar K_\mu \gamma^\mu & -\sqrt{\frac{-g_{tt}}{g_{zz}}} D_z -  m \sqrt{-g_{tt}}\\
\sqrt{\frac{-g_{tt}}{g_{zz}}} D_z -  m \sqrt{-g_{tt}} & i \bar K_\mu\gamma^\mu
\end{array}
\right)
\left(
\begin{array}{c}
\phi_+
\\
\phi_-
\end{array}
\right). \nonumber \\  
\end{eqnarray}
Here 
\begin{eqnarray}
\bar K_{\mu} \equiv \sqrt{\frac{-g_{tt}}{g_{ii}}} K_{\mu} = (-i (\partial_0 - i q A_0) \,, - i \sqrt{\frac{-g_{tt}}{g_{ii}}} (\partial_i - i q A_i))) \,,
\end{eqnarray}
and $D_z \equiv \partial_z - i q A_z $. 
In view of this, we consider the following matrix equation so that the above matrix is diagonalized,
\begin{eqnarray}
\left(
\begin{array}{cc}
0 & -\sqrt{\frac{-g_{tt}}{g_{zz}}} D_z - m \sqrt{-g_{tt}} \\
\sqrt{\frac{-g_{tt}}{g_{zz}}} D_z -  m  \sqrt{-g_{tt}} & 0
\end{array}
\right)
\left(
\begin{array}{c}
\phi_+
\\
\phi_-
\end{array}
\right)
= -m_l
\left(
\begin{array}{c}
\phi_+
\\
\phi_-
\end{array}
\right)
\end{eqnarray}
where $m_l$ is some eigenvalue. We write\footnote{The reason why we took a common factor $g_\pm(z)$ for the 2-spinor $\phi_\pm$ is
that this $z$-dependent factor go through the $\gamma$ matrices in $\bar K_\mu\gamma^\mu$ such that the $z$ integration
can be done independently as (\ref{aftergzgoesthrough1}), (\ref{aftergzgoesthrough2}).}
\begin{eqnarray}
\left(
\begin{array}{c}
\phi_+
\\
\phi_-
\end{array}
\right)
\equiv
\sum_l
\left(
\begin{array}{c}
g_+^{(l)}(z)\Psi^{(l)}(t,x,y)
\\
g_-^{(l)}(z)\Psi^{(l)}(t,x,y)
\end{array}
\right)
\label{phipms}
\end{eqnarray}
where $\Psi$ is a two-spinor which is a function of $(t,x,y)$, and $g_\pm(z)$ are scalar functions.
These $g_\pm(z)$ are required to satisfy
\begin{eqnarray}
\left(
\begin{array}{cc}
0 & -\sqrt{\frac{-g_{tt}}{g_{zz}}} D_z - \sqrt{-g_{tt}} m \\
\sqrt{\frac{-g_{tt}}{g_{zz}}} D_z - \sqrt{-g_{tt}} m & 0
\end{array}
\right)
\left(
\begin{array}{c}
g_+^{(l)}(z)
\\
g_-^{(l)}(z)
\end{array}
\right)
= -m_l
\left(
\begin{array}{c}
g_+^{(l)}(z)
\\
g_-^{(l)}(z)
\end{array}
\right). \quad \quad \quad
\end{eqnarray}

With explicit eigen wave functions, we can reduce the action to a 3-dimensional action.
Substituting \eqref{phipms} to
the action, we obtain
\begin{eqnarray}
S_{\rm fermion} =  
\label{aftergzgoesthrough1}
 \sum_{l,m} \int \! d^3xdz  \sqrt{\frac{g_{zz}}{-g_{tt}}}
\, i \, \bar{\Psi}^{(l)} (i \bar K_\mu\gamma^\mu - m_m)\Psi^{(m)}
\left(g_+^{(l)\, *}(z) g_+^{(m)}(z) + g_-^{(l) \, *}(z) g_-^{(m)}(z)\right) \,. \nonumber 
\\ 
\end{eqnarray}
The $z$ integration gives a normalization for the fermion which we require as
\begin{eqnarray}
\label{aftergzgoesthrough2}
\int \! dz \; \sqrt{\frac{g_{zz}}{-g_{tt}}} 
\left(g_+^{(l)\, *}(z) g_+^{(m)}(z) + g_-^{(l)\, *}(z) g_-^{(m)}(z)\right) = \delta_{lm}.
\end{eqnarray}
Furthermore, we calculate the chemical potential for each mode as
\begin{eqnarray}
\mu_{lm} \equiv 
\int \! dz \; \sqrt{\frac{g_{zz}}{-g_{tt}}} 
A_0(z)\left(g_+^{(l)\, *}(z) g_+^{(m)}(z)+ g_-^{(l)\, *}(z) g_-^{(m)}(z)\right).
\end{eqnarray}
Here we assume that this chemical potential is diagonal 
\begin{eqnarray}
\mu_{lm} = \mu_l \delta_{lm} \,,
\end{eqnarray}
and also consider only the case below where  
$g_{ii}(z) =-g_{tt}(z)$ such that 
dimensional reduction along $z$ ensures the Lorentz invariance in the $t$-$x$-$y$ spacetime\footnote{In general such diagonalization makes the diagonal matrix $m_l$ off-diagonal.  
So choosing the basis such that both $\mu_{lm}$ and $m_l$ diagonal 
is impossible for a generic $A_0(z)$.  
Similarly choosing $g_{tt}(z) = - g_{ii}(z)$ 
is impossible for a generic metric $g_{\mu\nu}$ such as the Lifshitz form. 
The argument in the previous subsection using the $1+0$-dimensional
picture does not refer to these assumptions.}. 
Then we finally obtain the decomposition
\begin{eqnarray}
\label{B133dimaction}
S_{\rm fermion} = \sum_l 
\int \! d^3x   \; i \, \bar{\Psi}^{(l)} (D_\mu\gamma^\mu - m_l)\Psi^{(l)},
\end{eqnarray}
where $D_i \equiv \partial_i - i q A_i$ and $D_0 \equiv \partial_0 - i q \mu_l$.

The on-shell condition for the $l$-th fermion is 
$E_l + q \mu_l = \sqrt{m_l^2 + k^2}$ where $k$ is the magnitude of the momentum in the $(x,y)$ space.\footnote{This $E_l$ coincides with $E_{l,n}$ given in (\ref{onedimdiraceq}), because 
both $E$'s, which is the eigenvalue of $i \partial_t$, are obtained from the equations of motion of same action; One is from 4-dimensional viewpoint (\ref{B54dimaction}), 
and the other is from dimensionally reduced 
3-dimensional viewpoint (\ref{B133dimaction}).}
The derivative in the Dirac operator has eigenvalues $\partial_\mu = (-i(w+q\mu_l), ik_1, ik_2)$.
In the presence of 
the magnetic field, the momentum $k$ is replaced by the Landau levels.

Let us proceed to calculate the free energy.
We bring the Tr Log into the following expression,
\begin{eqnarray}
{\cal F}_{\rm fermion}  
&\equiv&
\sum_l {\rm Tr} \;{\rm Log} (D_\mu \gamma^\mu - m_l)
= \sum_l {\rm Tr} \int^\infty_{m_l} \! dt \; {\rm Tr}\frac{1}{D_\mu \gamma^\mu -t}
\nonumber \\
&=&
\sum_l \int^\infty_{m_l} \! \! dt 
\int \!\frac{dw d^2k}{(2\pi)^3}
\frac{2t}{(w+q\mu_l)^2-k^2-t^2} \nonumber \\
&\equiv& \sum_l {\cal F}^l_{\rm fermion} \,.
\end{eqnarray}
As you can easily check, this integral is divergent. Therefore, we need to subtract the 
vacuum contribution (the free energy with $\mu_l=0$) which corresponds to the Dirac fermi sea. Then
\begin{eqnarray}
{\cal F}^{{\rm (reno)} \, l}_{\rm fermion} \equiv {\cal F}^l_{\rm fermion} (\mu_l) - {\cal F}^l_{\rm fermion}(\mu_l=0) \,.
\end{eqnarray}
Now we perform the off-shell $w$ integration. 
The standard path in the complex $w$-plane
for the
integration of $w$, which rounds the upper half plane, concerns two 
poles at $w=-q\mu_l \pm \sqrt{k^2 + t^2}$. In the $i\epsilon$ prescription,
the only relevant pole in the path integral is the one on the negative real 
axis of $w$. However, note that now we have a contribution from the chemical
potential, thus there is the case when the both of the two poles are on the
negative real axis. 
\begin{eqnarray}
{\cal F}^l_{\rm fermion}(\mu_l)
&=&
\int^\infty_{m_l} \! \! dt 
\int \!\frac{dw d^2k}{(2\pi)^3}
\frac{2t}{(w+q\mu_l-\sqrt{k^2+t^2})(w+q\mu_l +\sqrt{k^2+t^2})}
\nonumber \\
& = &
\int^\infty_{m_l} \! \! dt 
\int \!\frac{d^2k}{(2\pi)^3}
\left[
\frac{2\pi i \; 2t}{-2\sqrt{k^2+t^2}}\theta(q\mu_l + \sqrt{k^2+t^2})
\right.
\nonumber \\
& & 
\left.\hspace{40mm}
+ \frac{2\pi i \; 2t}{2\sqrt{k^2+t^2}}\theta(q\mu_l-\sqrt{k^2+t^2})
\right].
\label{intw}
\end{eqnarray}
Note that the $\theta$ function in the first term is always equal to the unity. So, the first term is
independent of $\mu_l$.
The second term vanishes when $\mu_l=0$. Therefore, we find that ${\cal F}_{\rm fermion}^{\rm reno} = \sum_l {\cal F}^{{\rm (reno)} \, l}_{\rm fermion} $ 
coincides with the sum of the second term in \eqref{intw},
\begin{eqnarray}
{\cal F}_{\rm fermion}^{\rm reno} =
\sum_l {\rm tr} \int\frac{d^2k}{(2\pi)^3} \int^\infty_{m_l} \! dt \;\; 2\pi i \frac{2t}{2\sqrt{k^2+t^2}} \theta(q\mu_l-\sqrt{k^2+t^2}).
\label{fermireno}
\end{eqnarray}
The $\theta$ function represents the Fermi surface, since 
one satisfies the momentum constraint relevant to the chemical potential, $q\mu_l > \sqrt{k^2+t^2}$ 
(roughly speaking, $t$ is the mass of the
$l$-th mode of the fermion). In fact, this is a widely known technique 
\cite{Campbell:1974qu}.
Denoting the value of $t$ satisfying $q\mu_l-\sqrt{k^2+t^2}=0$ as $t_{*l}$,
\eqref{fermireno} is  integrated to
\begin{eqnarray}
{\cal F}_{\rm fermion}^{\rm reno} 
&=&\sum_l {\rm tr} \int\frac{d^2k}{(2\pi)^3} \int_{m_l}^{t_{*l}} \! dt \;\; 2\pi i \frac{2t}{2\sqrt{k^2+t^2}} 
\theta(t_{*l}-m_l)
\nonumber \\
&= &
\sum_l i \int \frac{d^2k}{(2\pi)^2}\left(\sqrt{k^2 + t_{*l}^2}-\sqrt{k^2+m_l^2}\right)\theta(t_{*l}-m_l)
 \nonumber \\
 &= &
\sum_l i \int \frac{d^2k}{(2\pi)^2}\left(q\mu_l-\sqrt{k^2+m_l^2}\right)\theta(t_{*l}-m_l)
 \nonumber \\
&=& -i \sum_l \int \frac{d^2k}{(2\pi)^2} E_l(k) \theta(-E_l(k)).
\end{eqnarray}
In the last equality, we have used the fact that $t_{*l}>m_l$ is equivalent to $\sqrt{k^2+m_l^2}-q\mu_l<0$.

In the presence of the magnetic field, the momentum $k^2$ is replaced by the Landau levels $2qBn$ 
with a non-negative integer $n$. The momentum integral is accordingly normalized as 
$dk_1dk_2 =2 \pi k dk = \pi d(k^2) = 2\pi qB dn$, so  
\begin{eqnarray}
{\cal F}^{({\rm reno})}_{\rm fermion} &= &
 -i \sum_l \; \frac{qB}{2\pi} \!\sum_n  E_{l,n} \theta(-E_{l,n}).
\end{eqnarray}
This expression is \eqref{fermione} in the Euclidean notation.

%%%%%%%%%%%%%%%%%%%%%%%%%%%%%%%%%%%%%%%%%%

\end{document}